\documentstyle[12pt]{article}
\newcommand{\gtmu}{\gamma_{\mu}}
\newcommand{\gtnu}{\gamma_{\nu}}
\newcommand{\gta}{\gamma_{\alpha}}

\newcommand{\gtl}{\gamma_{\lambda}}

\newcommand{\ghmu}{\gamma^{\mu}}

\newcommand{\gha}{\gamma^{\alpha}}

\begin{document}
  \newcommand{\EQN}{\label}              %
  \newcommand{\beq}{\begin{equation}}    %
  \newcommand{\eeq}{\end{equation}}      %
  \newcommand{\r}[1]{(\ref{#1})}         %
                                         %
                                         %
%%%%%%%%%%%%%%%%%%%%%%%%%%%%%%%%%%%%%%%%
\newcommand{\be}{\begin{equation}}
\newcommand{\ee}{\end{equation}}
\newcommand{\bea}{\begin{eqnarray}}
\newcommand{\eea}{\end{eqnarray}}
\newcommand{\beas}{\begin{eqnarray*}}
\newcommand{\eeas}{\end{eqnarray*}}
\newcommand{\bdm}{\begin{displaymath}}
\newcommand{\edm}{\end{displaymath}}
\newcommand{\ba}{\begin{array}}
\newcommand{\ea}{\end{array}}
\newcommand{\bi}{\begin{itemize}}
\newcommand{\ei}{\end{itemize}}
\newcommand{\ben}{\begin{enumerate}}
\newcommand{\een}{\end{enumerate}}
\newcommand{\bd}{\begin{description}}
\newcommand{\ed}{\end{description}}
\newcommand{\bq}{\begin{quote}}
\newcommand{\eq}{\end{quote}}
\newcommand{\bfg}{\begin{figure}}
\newcommand{\efg}{\end{figure}}
\newcommand{\bt}{\begin{table}}
\newcommand{\et}{\end{table}}
\newcommand{\btb}{\begin{tabular}}
\newcommand{\etb}{\end{tabular}}

\newcommand{\g}{\gamma}
\newcommand{\GFt}{$\gamma_{5} \ $}
\newcommand{\GFm}{\gamma_{5}}
\newcommand{\GDt}{$\Gamma^{\lbrack 3 \rbrack}$}
\newcommand{\GDm}{\Gamma^{\lbrack 3 \rbrack}}
\newcommand{\GNt}{$\Gamma^{\lbrack n \rbrack}$}
\newcommand{\GNm}{\Gamma^{\lbrack n \rbrack}}
\newcommand{\calO}{${\cal O}$}
\newcommand{\calOp}{${\cal O'}$}
\newcommand{\calOm}{{\cal O}}
\newcommand{\calOpm}{{\cal O'}}
\newcommand{\eue}{\frac{1}{\epsilon}}
\newcommand{\LMM}{\ln(\frac{\mu^2}{m^2})}
\newcommand{\unit}{1\hspace{-0.36em}1}
%\everymath={\displaystyle}
\newcommand{\nc}{\newcommand}
%%%%this is two new Moscow's group command %%%%%%%%
\nc{\renc}{\renewcommand}                         %
\nc{\ice}[1]{\relax}                              %
\nc{\wtd}{\widetilde}
\nc{\dsp}{\displaystyle}
\begin{titlepage}
\noindent
\hfill TTP93--26 \\
\mbox{}
\hfill November 1993 \\
%
% Titel
%
\vspace{0.5cm}
\begin{center}
  \begin{Large}
    \begin{bf}
        Renormalization of 4--Quark Operators\\[0.4cm]
        and QCD--Sum Rules  \\
    \end{bf}
  \end{Large}
%
% Author
%
 \vspace{1.2cm}
 \begin{large}
L.-E.Adam   \\[3mm]
    Institut f\"ur Theoretische Teilchenphysik\\
    Universit\"at Karlsruhe\\
    Kaiserstr. 12,    Postfach 6980\\ %[2mm]
    76128 Karlsruhe, Germany\\ [5mm]
K.G.Chetyrkin  \\[3mm]
    Institute for Nuclear Research\\
    Russian Academy of Sciences\\
    60th October Anniversary Prospect 7a\\
    Moscow, 117312, Russia \\
 \end{large}
%
% Abstract
%
  \vspace{3cm}
  {\bf Abstract}
\end{center}
\begin{quotation}
\noindent
We compute the renormalization mismatch  displayed in 1--loop
approximation by classically equivalent 4-quark operators  and
coming from different possible definitions of the $\gamma_5$ matrix
in  dimensional regularization. The result is then employed to study
the effect of the various treatments of  $\gamma_5$ upon the size of
radiative corrections to 4-quark condensates in the QCD sum rules
for $\rho$ and $A_1$ mesons. We find that a fully anticommuting
$\gamma_5$ which automatically respects non-anomalous chiral
Ward-Slavnov identities leads to considerably smaller corrections
and reduces theoretical uncertainty in the QCD prediction for the
$\tau$ hadronic decay rate.
\end{quotation}
\end{titlepage}

\newpage

\section{Introduction}

Dimensional regularization (DR) \cite{dim.reg} coupled with the
minimal subtraction scheme (MS) \cite{ms} is widely acknowledged as
theoretically sound and  extremely valuable in practice calculational
framework.  In a theory like QCD with parity conserving amplitudes it
respects all the symmetries of the theory such as gauge invariance,
Bose symmetry and Ward-Slavnov identities. In addition,
dimensionaly regularized divergent Feynman integrals are in many
respects very similar to the convergent ones, which allowed to
develop quite powerful methods of their  analytical calculation.

A notorious shortcoming of DR is the absence of a natural
generalization of the Dirac matrix \GFt  \  to noninteger values
$D\not=4$ of the dimension of space-time.  Indeed, the well-known
relation
\beq
  {\rm tr}(\GFm \g_{\mu_1}\g_{\mu_2}\g_{\mu_3}\g_{\mu_4})
  = - 4i\epsilon_{\mu_1\mu_2\mu_3\mu_4}
  \EQN{rel1}
\eeq
which in fact fixes \GFt  \  in 4 dimensions, is not compatible
with another property of \GFt  \  valid at $D=4$
\beq
  \{ \GFm,\gamma_\mu \} =0, %\ \ \  \mu= 1 \dots D
  \EQN{rel2}
\eeq
The problem is  that the totally antisymmetric $\epsilon$-tensor is
a purely 4-dimensional object and thus can not be self-consistently
continued to $D$  dimensions. It may be shown that
if \r{rel2} is to be held literally for  all values of $\mu= 1 \dots D$
(and the standard properties of the trace operation like cyclycity
is respected) then  the rhs of \r{rel1} should vanish  identically
\cite{BM77a}.    Such a defined  \GFt  \  does not, hence, go
smoothly  at $D\to 4$ to the 4-dimensional
\beq
 \GFm^{(4)} = - i\g^0\g^1\g^2\g^3
\EQN{gf4}
\eeq
in contradiction with general requirements to which  every
self-consistent regularization must meet \cite{Bonneau90}.  From the
same point of view the relation \r{rel2} may be freely  violated
for the values of index $\mu$ lying outside of the physical
4-dimensional subspace.

The existing solutions of the problem   may be divided into two
classes.  The first  one  favours a nonvanishing anticommutator
$\{\GFm,\gamma_\mu \}$ at $\mu \not=0,1,2,3$
and is discussed in general in \cite{gamma5mu} and \cite{Thompson85}.
An explicit version of such a scheme was proposed by {}'t Hooft and
Veltman and elaborated by Breitenlohner and Maison
\cite{dim.reg,BM77a}. Another class keeps \r{rel2} intact at least
for amplitudes without traces with odd numbers of \GFt  \  matrices
while trying to reproduce  the identity \r{rel1} either by hand
\cite{Chanowitz79} or, in a bit more sophisticated manner, by
redefining the trace operation for the traces with odd number of \GFt
\cite{Kreimer90,Korner92}.

Leaving aside a (rather academic) task  of finding an explicit formal
proof of self-consistency of these solutions, we will study in this
work another aspect of the \GFt  \ problem --- different
possibilities of choosing between classically equivalent 4--quark
operators which  begin to differ in higher orders due to different
treatment of the \GFt  \  matrix. As we shall see below, different
choices lead to different theoretical predictions of the QCD sum
rule method \cite{SVZ79} because of a commonly accepted approximate
procedure of the vacuum saturation in estimating the vacuum
expectation values of 4-quark operators.  On considering  a number
of particular physical problems, it will be demonstrated that the
choice of fully anticommuting \GFt considerably reduces theoretical
uncertainty coming from higher order corrections.

\section{The $\gamma_5$ %
ambiguity in 4-quark operators
%and its effect upon
%their vacuum expectation values
}

The problem under discussion appears even in cases when initial
amplitudes comprise no \GFt  \  matrix at all. Indeed, let us consider
the standard derivation of the QCD sum rules for the $\rho$
and $A_1$ mesons \cite{SVZ79}. It starts from constructing the  Wilson
expansion \cite{Wilson69} for the correlator
\beq
  \Pi_{\mu\nu}(q) = i \, \int dx \, e^{iqx}
               <Tj_{\mu}(x)j^\dagger_{\nu}(0)>\hspace{-0.5em}_{_0}
                   = (q_{\mu}q_{\nu} - g_{\mu\nu} q^2 ) \Pi(Q^2)
  \EQN{2.1}
\eeq
\beq
   \Pi(Q^2) = C_0 + \frac{C_4 <{\cal O}_4>
                    \hspace{-0.5em}_{_0}}{Q^4}
                  + \frac{C^{i,V/A}_6 <{\cal O}^i_6>
                    \hspace{-0.5em}_{_0}}{Q^6} + \ldots
  \EQN{2.2}
\eeq
where $j_{\mu} =  \bar u \ \gtmu(\GFm)  d $
is the current for the $\rho (A_1)$ meson.

The power corrections of order $O(1/Q^6)$  to  the rhs of the
expansion \r{2.2} come from 4-quark condensates (that is the
vacuum expectation values (VEV) of 4-quark operators) and lead
to most important numerically non-perturbative corrections to  the
sum rule for the $\rho$ meson. In the lowest order of
perturbation theory the contributions of the  4-quark operators
are well-known  and read \cite{SVZ79}
\beq
    \frac{C^{i,V}_6 Q^i_6}{g^2}
  = -\frac{2}{9}
      (\bar u\ \gtmu t^a u +\bar d\ \gtmu t^a d)\cdot
                      (\bar\Psi\  \ghmu t^a \Psi)
   -  {2}
       \bar u\ \GFm \gamma_\mu t^a d
    \, \bar d\ \GFm   \gamma_\mu t^a u
     \EQN{2.3a}
\eeq
for the vector correlator
and
\beq
    \frac{C^{i,A}_6 Q^i_6}{g^2}
  = -\frac{2}{9}
      (\bar u\ \gtmu t^a u +\bar d\ \gtmu t^a d)\cdot
                      (\bar\Psi\  \ghmu t^a \Psi)
   -  {2}
      \bar u \gamma_\mu  t^a  d
   \, \bar d  \gamma_\mu t^a  u
     \EQN{2.3b}
\eeq
for the axial vector one.
Here  $t^a,\  a=1,\dots 8 $ are the colour Gell-Mann matrixes
normalized as
$\ Tr(t^a t^b) = \frac{1}{2} \delta^{ab}$; $ (\bar \Psi
\Gamma \Psi) $
is a shorthand for $\sum_{f=u,d,s} (\bar \Psi_f \Gamma \Psi_f) $;
$g$ is the quark gluon coupling constant such that
$\alpha_s = g^2/4\pi$.

Let us  consider in detail the structure of the second term in
\r{2.3a}
proportional to the 4-quark  operator
\beq
\calOm'_1 =
      \bar u\ \GFm \gamma_\mu t^a d \,
      \bar d\ \GFm \gamma_\mu t^a u.
\EQN{2.4a}
\eeq
In fact,  the very calculation  produces a bit  different
operator
\beq
  \calOm_1 =
      \bar u\ \GDm t^a d \, \bar d\ \GDm t^a u, \ \
      \GDm \equiv \frac{1}{2} (\gtmu \gtnu \gtl - \gtl \gtnu \gtmu )
  \EQN{2.5a}
\eeq
while the result \r{2.3a} is obtained after a use of the identities
\beq
  \GDm \bigotimes \GDm =
  \,6 \GFm \gamma_\mu \bigotimes \GFm \gamma_\mu.
  \EQN{2.6}
\eeq
Similarly, the initial form of the second term in \r{2.3b}
\beq
\calOm'_2 =
      \bar u  \gamma_\mu t^a d \,\bar d \gamma_\mu t^a u.
\EQN{2.4b}
\eeq
is (with an appropriately changed coefficient function)
\beq
\calOm_1 =
      \bar u\ \GDm \GFm t^a d \, \bar d\ \GDm \GFm  t^a u,
\EQN{2.5b}
\eeq

The relation \r{2.6} is valid in 4-dimensional space due to the
very definition of the $\GFm^{(4)}$ matrix \r{gf4}.  However,  it
loses any operational sence at generic $D\not=4$ within the framework
of dimensional regularization.  This implies that if both
operators $\calOm_1$ and $\calOm'_1$  are to be minimally renormalized, then
the
difference
\beq \delta \calOm_1 = \calOm_1 - 6\calOm'_1  = {\cal O}(\alpha_s)
\EQN{2.7}
\eeq
is an {\em evanescent} operator \cite{Collins84} which vanishes in
the tree approximation but typically receives nonzero contributions
in higher orders.  This also  means that higher order corrections to
the coefficient functions (CF) $C_6^i$  also depend on specifying
which operators are chosen as basic  ones.  It should be stressed
that this $\GFm$ ambiguity  comes from the absence of a  unique
canonical representation of 4-quark operators in dimensional
regularization, which, in turn, does not allow to define the idea of
minimal subtraction of UV poles in an unambiguous way.

The above discussion does not imply in any means that  the
theoretical predictions for the contributions to the rhs of \r{2.2}
from  4-quark operators are ill-defined in principle: the VEV's of
the latter should, of course, also depend on the choice of the
renormalization prescription in such a way that  the whole
combination $\sum_i C^i_6 \calOm^i_6$  is  invariant   as it must be.

The {\em real}
problem lies  in our present inability to find the VEV's of
4-quark operators in a sufficiently accurate way.
Indeed, the common practice is to use the vacuum
saturation (VS)  procedure \cite{SVZ79}. It   states that
the VEV of a 4-quark operator
${\cal O} = (\bar \Psi \Gamma_1 \Psi) (\bar \Psi \Gamma_2 \Psi) $
may be estimated according to the following formula
\beq
   <{\cal O}> \hspace{-0.5em}_{_0} = \frac{1}{144}\, \lbrack Tr(\Gamma_1)
       Tr(\Gamma_2) - Tr(\Gamma_1 \Gamma_2) \rbrack \,\cdot <\bar qq>^2
   \EQN{vd}
\eeq
where $\bar qq = \bar uu, \bar dd$ or $ \bar ss $.

It is clear that  this procedure is too rough and does not feel in
any way such fine details as the concrete mode of the renormalization
of a 4-quark operator.  Hence, if one would like  to estimate the
effect of higher order corrections to the CF's $C_6^i$ the numerical
results will do depend on specifying 4-quark operators in $D$
dimensions. Unfortunately, these corrections prove to be of
considerable size (see  \cite{CheSpi88}  and below) at least in
the operator basis featuring objects like
$\GDm  \bigotimes
\GDm $.
Let us try  to search for  a distinguished operator basis for which
the VS would presumably work better.

Of course, from a general point of view there is an infinitely large
variety  of possible choices of a (different at $D\not=4$ and
equivalent at $D=4$)  4-quark operator. For instance, one
could mix operators $\calOm_1$ and $\calOm'_1$ as follows
\beq
  \calOm{''}_1   = \calOm_1\cdot sin^2 \alpha
                   +
                   6\,\calOm'_1\cdot cos^2 \alpha
  \EQN{2.8}
\eeq
with an arbitrary angle $\alpha \in \{-\pi/2, \pi/2\}$.

On the other hand  a lot of calculational
experience invariably demonstrates
that if  amplitudes under investigation do not comprise traces with
an odd number of \GFt  \  at all then the use of a fully
anticommuting \GFt  \  is quite self-consistent and leads to great
simplifications in doing practical calculations.  Much more
important,  this choice is in a sense unique for the amplitudes
from above described class (these will be referred to non-singlet ones
in the following) as it  respects  the chiral
$SU_L(3) \bigotimes  SU_R(3)$
symmetry of the QCD lagrangian.  This is the case since on the purely
diagrammatic level the chiral symmetry relies on the possibility of
freely anticommutating of  \GFt with $\gamma_\mu$ at generic value of
the index $\mu$.  It seems very reasonable to employ
identities like \r{2.6} in order to maximally simplify
kinematical structures appearing. Thus, we will
treat the operators
$\calOm'_1$ and $\calOm'_2$
with completely anticommutating \GFt as  ``natural'' ones and
reexpress through them the operators\footnote{Note that we are
concerned only with 4-quark operators appearing in the tree
approximation; for operators which show up first  in 1-looop
corrections the subtlety under discussion  becomes relevant starting
at two loops.}  $\calOm_1$ and $\calOm_2$.

A simple calculation
gives:
\beq
  \ba{l}
  \dsp
     {\cal O}_1
  %\rule[-4mm]{0mm}{10mm}
  = \     ( 1 -
        \frac{\alpha_s}{4\pi}\, \frac{22}{3})\cdot 6 \, {\cal O'}_1
      +  \frac{25\alpha_s}{4\pi}
         \bar u \ \gta t^a  d \, \bar d\ \gha t^a u\
\\
\dsp
  +      \frac{40}{3}\, \frac{\alpha_s}{4\pi}
         \bar u \ \gta  d \,  \bar d\ \gta  u\
+        \frac{2}{9}\, \frac{\alpha_s}{4\pi}
        (\bar u \ \gta t^a  u + \bar d\ \gha t^a d)\,
                                              (\bar \Psi \gha t^a \Psi)
,
\ea
\EQN{d1}
\eeq

\beq
\ba{l}
\dsp
   {\cal O}_2
%\rule[-4mm]{0mm}{10mm}
  = \     ( 1 -
        \frac{\alpha_s}{4\pi}\, \frac{22}{3})\cdot  6\, {\cal O'}_2
      +  \frac{25\alpha_s}{4\pi}
         \bar u \ \gta \GFm t^a  d \, \bar d\ \gha \GFm t^a u\
\\
\dsp
  +      \frac{40}{3}\, \frac{\alpha_s}{4\pi}
         \bar u \ \gta \GFm  d \,  \bar d\ \gta \GFm  u\
+        \frac{2}{9}\, \frac{\alpha_s}{4\pi}
        (\bar u \ \gta t^a  u + \bar d\ \gha t^a d)\,
                                                 (\bar \Psi \gha t^a \Psi).
\ea
\EQN{d2}
\eeq

To finish this section, a few technical words about the calculation.
It is clear that the  very difference $\calOm_i - \calOm'_i$  vanishes   in
the classical limit and consequently  may come  only from   UV
divergences which manifest themselves as poles in $\epsilon$ in DR
\cite{Collins84}. Thus, the calculation is conveniently  performed
hand in hand with the standard procedure of evaluating one-loop
anomalous dimensions of 4-quark operators
\protect\cite{SVZ79} (see also \cite{Buras89}, where similar
calculations were performed in the  context of evaluating
QCD corrections to   the effective weak Hamiltonian.

%\section{The $\gamma_5$ %
%ambiguity and the QCD sum rule for the $\rho$ meson}
\section{ QCD sum rules for the $\rho$ and $A_1$ mesons
        }

An  explicit calculation of one-loop
corrections to 4-quark condensates  in \r{2.3a}
gives  \cite{CheSpi88}
(in the commonly accepted $\overline{MS}$ scheme \cite{MSbar};
$L = \ln\frac{\mu^2}{Q^2}$ )
\bea
    \frac{C^{i,V}_6 Q^i_6}{g^2}
  &=& -\frac{2}{9}\,\left(1+\frac{\alpha_s}{\pi} \left\lbrack \frac{95}{72} L
                             + \frac{107}{48} \right\rbrack \right) \
      (\bar u\ \gtmu t^a u +\bar d\ \gtmu t^a d)\cdot (\bar\Psi\ \ghmu t^a\Psi)
     \label{S1}    \label{r18} \\
  & & -\frac{1}{3}\,\left(1+\frac{\alpha_s}{\pi}\left \lbrack \frac{9}{8} L
                             + \frac{431}{96} \right\rbrack \right) \
     \bar u\ \GDm t^a d \, \bar d\ \GDm t^a u
     \label{S2}    \label{r19}  \\
  & & -\frac{\alpha_s}{24 \pi} \,
      \left\{ \begin{array}{lll}
\rule[-4mm]{0mm}{10mm}        & ( 16 L - 12 )
       & \bar u \Gamma_{\mu} d \, \bar d \ \Gamma_{\mu} u               \\
\rule[-4mm]{0mm}{10mm}       +& (30 L - \frac{45}{2})
       & \bar u \Gamma_{\mu} t^a d\,  \bar d \ \Gamma_{\mu} t^a u      \\
\rule[-4mm]{0mm}{10mm}       +& ( \frac{16}{9} \, L - \frac{8}{27} )
       & (\bar \Psi \ \gtmu t^a \Psi )^2                                   \\
\rule[-4mm]{0mm}{10mm}       +& (\frac{16}{9} \, L + \frac{56}{27} )
       & (\bar u\ \gtmu \GFm  u + \bar d\ \gtmu \GFm  d)
        (\bar\Psi\ \ghmu \GFm  \Psi)                                       \\
\rule[-4mm]{0mm}{10mm}       +& (\frac{10}{3} \, L + \frac{35}{9} )
       & (\bar u\ \gtmu \GFm t^a u + \bar d\ \gtmu \GFm t^a d)
         (\bar\Psi\ \ghmu \GFm t^a \Psi)
      \end{array} \right\}
\label{3.1}
{}.
\eea
For the axial vector correlator the corrections look exactly the
same, only the matrices \GDt \ and
$\Gamma_{\mu}\,\,\,{}$ should be multiplied
by \GFt.
After the use of the VS approximation we get for the $\rho$ meson
\beq
   C^{i,V}_6<{\cal O}^i_6> \hspace{-0.5em}_{_0} = -\frac{224 \pi}{81}
   \left(
         1 + (\frac{705}{112}+ \frac{13}{252}L)\frac{\alpha_s}{\pi}
   \right)
   (\alpha_s <\bar qq>^2 \hspace{-0.5em}_{_0})
\EQN{3.2a}
\eeq
and for the $A_1$ meson
\beq
   C^{i,A}_6<{\cal O}^i_6> \hspace{-0.5em}_{_0} = \frac{352 \pi}{81}
\left(
1 + (\frac{777}{176} + \frac{149}{396}L)\frac{\alpha_s}{\pi}
\right)
   (\alpha_s <\bar qq>^2 \hspace{-0.5em}_{_0})
\EQN{3.3a}
\eeq
At last, a  straightforward application
of \r{d1},\r{d2} leads us to a
new result for  (\ref{r18}-\ref{3.1}) and consequently for that of
\r{3.2a}:
\beq
   C^{i,V}_6<{\cal O}^{i}_6> \hspace{-0.5em}_{_0} = -\frac{224 \pi}{81}
\left(
1 + (\frac{685}{336}+ \frac{13}{252}L)\frac{\alpha_s}{\pi}
\right)
   (\alpha_s <\bar qq>^2) \hspace{-0.5em}_{_0}
\EQN{3.2b}
\eeq
and for the $A_1$ meson
\beq
   C^{i,A}_6<{\cal O}^{i}_6> \hspace{-0.5em}_{_0} = \frac{352 \pi}{81}
   \left(
         1 + (\frac{917}{528} + \frac{149}{396}L)\frac{\alpha_s}{\pi}
   \right)
   (\alpha_s <\bar qq>^2 \hspace{-0.5em}_{_0})
\EQN{3.3b}
\eeq
Thus, we observe that uncomfortably large radiative corrections in
\r{3.2a}and \r{3.3a} are transformed
into  quite moderate ones after the above described
transition to the new basis of 4-quark operators. Moreover, now
the $\alpha_s$ corrections get of approximately the same (relative)
size for the vector and axial correlators  --- a fact that will be of
special importance in the next section.

Note also that the terms proportional to $L$
in \r{3.2b} and \r{3.3b} are quite small.
To our opinion,   this fact demonstrates that
the effects due to  two-loop  anomalous dimensions of operators
in \r{3.1} are presumably small and may be  neglected.
On the other hand,  the combination of operators
$\alpha_s(\mu)  <\bar qq>(\mu)   <\bar qq>(\mu)$
has only very weak dependence on $\mu$ as its anomalous dimensions
is fortunately rather small. This means that if the VS
approximation is a good one at some scale $\mu$, then it should
be considered as a reasonable approximation at
a diffrent scale $\mu'$.

\section{The tau hadronic width
        }

As was pointed out on \cite{SchTra84,Bra88,NarPic88}
some time ago,  the methods of perturbative QCD
can be applied to estimate the decay rate ratio
\beq
R_{\tau} = \frac{\Gamma(\tau\rightarrow\nu_{\tau} {\rm hadrons})}
           {\Gamma(\tau\rightarrow\nu_{\tau}e \bar{\nu}_e)}.
\EQN{4.1}
\eeq
%Further work has been dealing  with  non-perturbative effects
%\cite{SchTra84,NarPic88}
%as well as with  electroweak corrections \cite{MarSir88,BraLi90} to this
%quantity.
An updated theoretical discussion of  the ratio
\r{4.1} within the QCD frameworks  was recently given in
\cite{BraNarPic91}.

Schematically,   $R_\tau$   may  represented in the following form
(in the chiral limit of the massless $u$, $d$ and $s$ quarks and
neglecting the logarithmic dependence of the Wilson coefficients)
\beq
R_\tau =    R^0_\tau
\left\{1+\frac{\alpha_s}{\pi}
+5.202(\frac{\alpha_s}{\pi})^2 +26.37
(\frac{\alpha_s}{\pi})^3
+ \delta^{D =6}
+ \dots
\right\}.
\EQN{4.2}
\eeq
Here  the perturbative coefficients are appropriate for the number of
flavours $f=3$ and $\alpha_s =  \alpha_s(M_\tau)$
and  the power correction of order $1/M_\tau^6$
is directly related to the contribution of 4-quark operators to
\r{2.2}, viz.
\beq
\delta^{D =6} = (\delta^{D =6}_V + \delta^{D = 6}_A)/2, \ \
\delta^{D =6}_{V/A}
=
24\pi^2 \frac{\dsp -\sum_i
( C^{i,V/A}_{6}) < \calOm >_i )}%
{M_\tau^6}
\EQN{4.3}
\eeq
As was argued  in \cite{BraNarPic91} the only significant
sources of  uncertainty in the QCD prediction for $R_\tau$
are due to the uncalculated perturbative QCD correction of order
$\alpha_s^4(M_\tau)$ and  due to the  power correction of order
$1/M_\tau^6$. In this section we will concentrate on the latter.

The estimation of the power correction  made in
\cite{BraNarPic91} amounts essentially to keeping the lowest order
contribution to the CF $\delta^{D =6}$ and using  the VS
approximation. The result is
\beq
\delta^{D =6}_{V/A} =
\left(
      \ba{c}
        7
        \\
       -11
       \ea
\right)
\frac{256\pi^3}{27}
\frac{\rho \alpha_s <\bar{\psi}\psi>^2}{M_\tau^6}.
\EQN{4.4}
\eeq
Here
$
\rho \alpha_s <\bar{\psi}\psi>^2
\approx (3.8 \pm 2.0)\times
10^{-4} GeV^6
$ is an effective scale
invariant matrix element which is determined phenomenologically
\cite{rho1,rho2,rho3}. It is introduced in order to partially take
into account deviations from the VS approximation. An important
observation is that when the vector and  axial vector terms are
averaged to give $\delta^{D=6}$, there is a large mutual cancellation
between these contributions. It is the  assumption ---
that this cancellation also holds for the
uncertainties of separate
contributions\footnote{This would become true if
the prescription  \r{vd} could be made exact by
multiplying its rhs with  an (operator independent!)   correction
factor $\rho$.}
 ---,  which is   crucial in producing a rather  accurate   prediction
for $\alpha_s$ given in    \cite{BraNarPic91}, viz.
\beq
\alpha_s(M_\tau) = 0.34 \pm 0.004 .
\EQN{4.5}
\eeq
This assumption is certainly quite optimistic  and has
already been criticized in  \cite{Altarelli92,Chetyrkin93}.
{}From   formal point of view it is self-consistent {\em
provided}  one  may neglect higher order corrections to the
coefficient functions involved.  However, the above discussion
of the leading 1-loop corrections to the CF
$C_6^{i,V/A}$ clearly demonstrates
that  this neglection is very questionable   unless  one  deals with
the primed basis of 4-quark operators and avoids objects like
$\Gamma^{[3]}\bigotimes \Gamma^{[3]}$ in favour of those
featuring the completely anticommuting   $\GFm$.

\section{Conclusions}

In this work we have studied the problem of the dependence of
radiative corrections to the coefficient functions of 4-quark
condensates on the exact way of treating the $\gamma_5$ matrix and
related objects specifying a 4-quark operator.  It was argued  that
a choice of a fully anticommutating $\gamma_5$ matrix which respects
the chiral symmetry of the QCD lagrangian is theoretically preferred
and leads to better convergent perturbation series.
It also gives an extra argument in favour of the stability
of \r{4.4} and \r{4.5} with respect to radiative corrections.

We  would like to thank J. H. K\"uhn and A. Kwiatkowski for their
support and helpful discussions.

%\small

\end{document}